\documentclass{eas}
\usepackage{graphicx}
\begin{document}

%%-----------------------------
%%      the top matter
%%------------------------
\title{Radial mixing due to spiral--bar resonance overlap: Implications to the 
Milky Way} 
\runningtitle{Radial mixing due to resonance overlap}
\author{I. Minchev}\address{Observatoire Astronomique, Universit\'e de
Strasbourg, CNRS UMR 7550,
67000 Strasbourg, France}
\author{B. Famaey}\sameaddress{1}
\begin{abstract}
We have recently identified a previously unknown radial migration
mechanism resulting from the overlap of spiral and bar resonances in
galactic discs (Minchev \& Famaey \cite{mf10}, Minchev {\em et al.\/} \cite{mig2}).
This new mechanism is much more efficient than mixing by transient
spirals and its presence is unavoidable in all barred galaxies, such as 
our own Milky Way. The consequences of this are a strong flattening in the
metallicity gradient in the disc, an extended disc profile, and the
formation of a thick disc component, all taking place in only a couple
of Gyr. This timescale is drastically shorter than previously expected
and thus can put strong constraints on the longevity, strength and
pattern speeds of the Galactic bar and Spiral Structure. 
\end{abstract}
\maketitle
%%-----------------------------
%%      your text
%%-----------------------------
\section{Introduction}

In the last decades discrepancies in the solar neighborhood age-metallicity
relation have conclusively demonstrated that effective radial migration (i.e., 
redistribution of angular momentum) must be taking place in the Milky Way disc 
(e.g., Haywood \cite{haywood08}, Schonrich \& Binney \cite{schonrich09}). 
Until recently it was accepted that such mixing was 
solely caused by transient spirals (Sellwood \& Binney \cite{sellwood02}). 
However, Quillen {\em et al.\/} (\cite{quillen09}) showed that small 
satellites on radial, in-plane orbits can cause mixing in the outer disc and thus
account for the fraction of low-metallicity stars present in the solar
neighborhood Haywood (\cite{haywood08}). Note that the Milky Way (MW), as well 
as more than 2/3 of disc galaxies contain central bars. Are bars important for 
the process of radial migration?

\section{Resonance overlap of multiple patterns}

We have recently shown 
(Minchev \& Famaey \cite{mf10}, hereafter MF10) that a strong exchange of angular 
momentum occurs when a stellar disc is perturbed by a central bar and spiral 
structure (SS) simultaneously (see Fig. 1). By using test-particle simulations, 
we attributed this effect to the overlap or first and second order resonances of
each perturber. Both bars and spirals can participate to this overlap of resonances. 
The efficiency of this mechanism was later 
confirmed in fully self-consistent, Tree-SPH simulations, as well as high-resolution 
pure N-body simulations Minchev {\em et al.\/} (\cite{mig2}) (see Fig. 2).

\begin{figure}
\includegraphics[width=12cm]{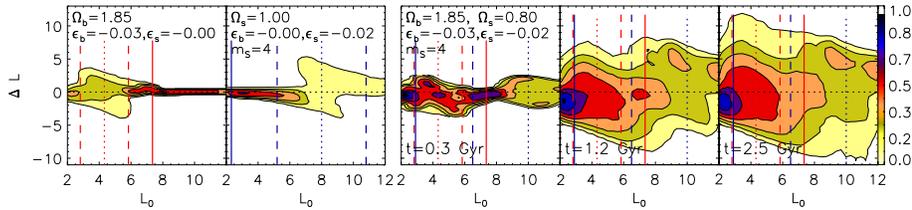}
\caption{Changes in angular momentum, $\Delta L$, as a function of the initial 
angular momentum, $L_0$. From left to right the first 2 panels show the effect of
a bar or a SS only, respectively, with parameters consistent with the MW. The 
simultaneous propagation of the same 
perturbers is shown in the following 3 panels for $t=0.3-2.5$~Gyr. The dotted 
lines show the corotation radii. The 2:1 and 4:1 LRs are indicated by the solid 
and dashed lines respectively (bar=red, spiral=blue). Figure is from MF10.}
\end{figure}

\section{Implications for the Milky Way}
\label{sec:mw}

To illustrate the effect of this migration mechanism in the MW we use a bar pattern 
speed of $\Omega_b=1.85\Omega_0$, where $\Omega_0$ is the angular velocity at the 
solar circle $r_0$, consistent with recent pattern speed estimates (Minchev 
{\em et al.\/} \cite{mnq07, minchev10}; Dehnen \cite{dehnen00}). For the bar
amplitude we use a value derived by Rodriguez-Fernandez \& Combes (\cite{rodriguez08}).
The MW SS parameters are much more uncertain. A detailed discussion can be found
in MF10 for the motivation of our choice of parameters.

The last 3 panels of Fig. 1 show the resonance overlap induced migration for a 
bar and SS consistent with the MW. As one can see, since this is a nonlinear 
effect, depending on the perturbers' amplitudes, strong mixing could occur in a 
short period of time. While there is no doubt that the MW has been 
affected by this process, it may be difficult to estimate exactly how. This 
migration mechanism is a strong function of the strengths of the MW bar and
SS. However, it may be incorrect to use the currently observed spiral and bar
amplitudes since most likely they have not been the same throughout the lifetime 
of the Galaxy. Since resonance overlap also induces stellar heating
(Minchev \& Quillen \cite{mq06}), we can put constraints on the amount 
of radial migration that has taken place in the MW is by requiring that we do not 
overheat the Galactic disc.

\begin{figure}
\includegraphics[width=12cm]{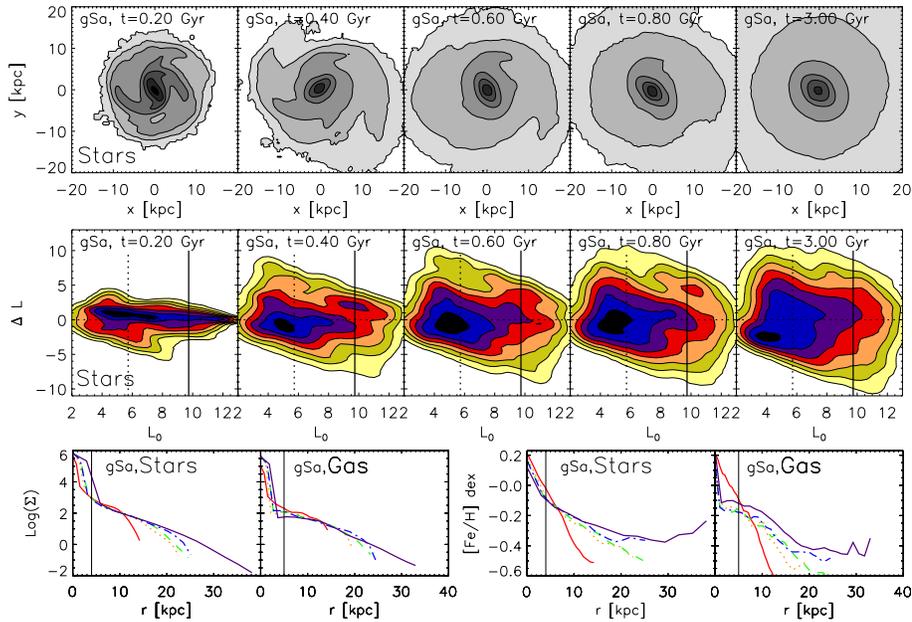}
\caption{
Results of a Tree-SPH simulation, studying the exchange of angular momentum due
to resonance overlap of bar and SS. {\bf Top row:} Time development of the
stellar disc density contours of a giant Sa galaxy. {\bf Second row:} $\Delta L$ as 
a function of the initial angular momentum, $L_0$. {\bf Bottom row:} The evolution 
of the radial 
profiles of surface density (left) and metallicity (right) for the stellar and 
gaseous discs. The initial disc scale-lengths are
indicated by the solid lines. The 5 time steps shown are as in the top row,
indicated by solid red, dotted orange, dashed green, dotted-dash blue and solid 
purple, respectively, from Minchev {\em et al.\/} (\cite{mig2}).
}
\end{figure}

\section{Conclusions}

Radial mixing caused by the resonance overlap of multiple patterns could be up to an
order of magnitude more effective than the transient spirals mechanism. This effect 
is non-linear, strongly dependent on the strengths of the perturbers. The signature 
of this mechanism is a bimodality in the changes of angular momentum in the disc 
with maxima near the bar's corotation and its outer Lindblad resonance (Figs. 1 
and 2). This is true regardless of the spiral pattern speed. This migration mechanism 
can create extended discs in both MW-mass (Fig. 2) and low-mass galaxies, such as 
NGC 300 and M33 (Minchev {\em et al.\/} \cite{mig2}, Fig. 4). It can also be responsible 
for the formation of a thick disc component early on in the galaxy evolution 
(Minchev {\em et al.\/} 2010, in preparation).

For bar and spiral parameters consistent with MW observations we find
that it takes $\sim$~3~Gyr to achieve the mixing for which transients require
9~Gyr (cf., MF10). In addition to radial mixing, spiral-bar coupling can account for 
the age-velocity relation (AVR) observed in the solar neighborhood. Note however that
the estimates for the MW spiral and bar strengths derived from observations 
{\it now} may simply be irrelevant for our understanding of the current state of 
mixing in the Galactic disc.

Ongoing and planned large Galactic surveys, such as RAVE, SEGUE, SIM Lite, GYES, 
LAMOST and GAIA, can search for signatures of the mechanism (cf, Minchev \&
Quillen \cite{mq08}). However, the most promising technique to put constraints on 
this mechanism in the MW is "chemical tagging" (Freeman \& Bland-Hawthorn 
\cite{freeman02}) which will become possible with the forthcoming spectroscopic 
survey HERMES. Therefore, in the next couple of years we may be in a position to
directly measure the spread of open clusters across the Galaxy as a function
of age, thus providing direct constraints on the amount of mixing in the MW
(Bland-Hawthorn {\em et al.\/} \cite{joss10}).

\acknowledgements
We are grateful to the organizers for a very interesting conference and a traveling 
grant. Special thanks to Catherine Turon.

%%-----------------------------
%%      your bibliography
%%-----------------------------

\end{document}